\documentclass[12pt,draft]{article}

\usepackage{amssymb}
\usepackage{latexsym}
\input psfig.sty

\newtheorem{theorem}{Theorem}[section]
\newtheorem{definition}[theorem]{Definition}
\newtheorem{proposition}[theorem]{Proposition}
\newtheorem{conjecture}[theorem]{Conjecture}

\newcommand{\nc}{\newcommand}
\nc{\N}{{\mathbb N}}
\nc{\R}{{\mathbb R}}

\begin{document}

\title{Note on a reformulation of the strong \\
cosmic censor conjecture based on computability}

\author{ 
G\'abor Etesi
\\ {\it Alfr\'ed R\'enyi Institute of Mathematics,}
\\ {\it  Hungarian Academy of Science,}
\\{\it Re\'altanoda u. 13-15. Budapest, H-1053 Hungary}
\\ {\tt etesi@math-inst.hu}}

\maketitle

\pagestyle{myheadings}
\markright{G. Etesi,: Cosmic censorship and computability}

\thispagestyle{empty}

\begin{abstract}
In this letter we provide a reformulation of the strong cosmic censor
conjecture taking into account recent results on Malament--Hogarth
space-times.

We claim that the strong version of the cosmic censor conjecture can be
formulated by postulating that a physically reasonable
space-time is either globally hyperbolic or possesses the
Malament--Hogarth property. But it is known that a Malament--Hogarth
space-time in principle is capable for performing non-Turing computations 
such as checking consistency of ZFC set theory.

In this way we get an intimate conjectured link between the cosmic
censorship scenario and computability theory.
\end{abstract}
\vspace{0.1in}

\centerline{PACS numbers: 04.20.Dw, 89.20.Ff}
\centerline{Keywords: {\it strong cosmic censorship, Malament--Hogarth
space-times, non-Turing computability}}

\section{Introduction}

There is a remarkable recent interest in the physical foundations of
computability theory and the Church--Turing Thesis. It turned out 
that algorithm and complexity theory, previously considered as very pure
mathematical subjects, have a deep link with basic concepts of physics.
At one hand now we can see that our deep and apparently pure mathematical
notion of a Turing machine involves indirect preconceptions
on space, time, motion and measurement. Hence it is straightforward to ask
whether different choices of physical theories for modeling these things
have some effect on our notions of computability or not. At the recent
stage of affairs it seems there are striking changes on the whole
structure of complexity and even computability theory if we move from
classical physics to quantum or relativistic theories. Even variants of
the Church--Turing Thesis cease to be valid in certain cases.  

For instance, by taking quantum mechanics as our background theory, Calude
and Pavlov have claimed in their recent paper that the famous Chaitin's
Omega number, a typical non-computable real number, is enumerable via an
advanced quantum computer \cite{cal-pav} while Kieu has recently proposed
an adiabatic quantum algorithm to attack Hilbert's tenth problem
\cite{kie}. 

In the same fashion if we use general relativity theory, powerful
gravitational computers can be built up which are also capable to break
Turing's barrier. Malament and Hogarth proposed a class of space-times,
now called as Malament--Hogarth space-times, admitting gravitational
computers for non-Turing computations \cite{hog1} \cite{hog2}. Hogarth'
construction uses anti-de Sitter space-time which is in the focus of
recent investigations in high energy physics. In the same spirit the
author and N\'emeti have constructed another example by exploiting
properties of the Kerr--Newman space-time \cite{ete-nem}. This space-time
is also relevant as the only possible final state of a collapsed,
massive, slowly rotating star of small electric charge. A general
introduction to the field is Chapter 4 of Earman's book
\cite{ear}.

On the other hand it is conjectured that these generalized computational
methods are not significant from a computational viewpoint only but,
in the case of quantum computers at least, they are also in connection
with our most fundamental physical concepts such as the standard model and
string theory \cite{bla} \cite{sch}.

The natural question arises if the same is true for gravitational
computers i.e., is there any pure physical characterization of the above
mentioned Malament--Hogarth space-times? In this short paper we try to
argue that these space-times also appear naturally in the strong cosmic
censorship scenario. Namely we claim that space-times
possessing powerful gravitational computers form the unstable borderline
separating the allowed and not-allowed space-times by the strong cosmic
censor conjecture (but these space-times are still considered as
``physically relevant'' i.e., are not ruled out by the strong cosmic
censor).

If our considerations are correct then we can establish a hidden link
betwix non-Turing computability and the most exciting open problem
of classical general relativity.

\section{Malament--Hogarth space-times}
In this section we introduce the concept of a Malament--Hogarth
space-time. As a motivation we mention that in these space-times, at least
in theory, one can construct powerful gravitational computers 
capable for computations beyond the Turing barrier. A typical example for
such a computation is checking of consistency of ZFC set theory. 

Then we prove a basic property of Malament--Hogarth space-times namely
they lack global hyperbolicity and distinguish the two main subclasses of
them. Finally we provide three physically relevant examples
possessing the Malament--Hogarth property.

Remember that the length of a non-spacelike, once continuously
differentiable curve $\gamma :\R\rightarrow N$ in a pseudo-Riemannian
manifold $(N,h)$ is the integral
\[\Vert\gamma\Vert =\int\limits_\gamma{\rm d}\gamma =\int\limits_\R
\sqrt{ -h(\dot\gamma (t),\dot\gamma (t))}\:{\rm
d}t=\int\limits_\R\sqrt{-\vert\dot\gamma (t)\vert^2_h}\:{\rm d}t\]
if exists. If this integral is unbounded we shall write $\Vert\gamma\Vert
=\infty$. Furthermore we will be using the following standard terminology.
Let $(M,g)$ be a space-time which is a solution to Einstein's equation
with a matter source represented by a stress-energy tensor $T$ obeying
the dominant energy condition on $M$. We will suppose this matter is {\it
fundamental} in the sense that the associated Einstein's equation (derived
as a variation of $T$ with respect to the metric $g$) can be put into the
form of a quasilinear, diagonal, second order, hyperbolic system of
partial differential equations. It is well-known that in this case $(M,g)$
admits a well-posed initial value formulation $(S,h,k)$ \cite{haw-ell}
\cite{wal}. Here $S$ is a connected, spacelike hypersurface, $h=g\vert_S$
is the restricted Riemannian metric on $S$ while $k$ is the second
fundamental form of $(S,h)$ as embedded in $(M,g)$. 

Now consider the following class of
space-times (cf. \cite{ete-nem} but also \cite{hog1}\cite{hog2}):
\begin{definition} Let $(S,h,k)$ be an initial data set for Einstein's
equation, with $(S,h)$ a complete Riemannian manifold. Suppose a
fundamental matter field is given represented by its stress-energy 
tensor $T$ satisfying the dominant energy condition. Let $(M,g)$ be a
maximal analytical extension (if exists) of the unique maximal Cauchy
development of the above initial data set.

Then $(M,g)$ is called a {\em Malament--Hogarth space-time} if there is a
future-directed timelike half-curve $\gamma_C :\R^+\rightarrow M$ such
that $\Vert\gamma_C\Vert =\infty$ and there is a point $p\in M$ satisfying
${\rm im}\gamma_C\subset J^-(p)$. The event $p\in M$ is called a {\em
Malament--Hogarth event}. $\Diamond$
\end{definition}
Note that if $(M,g)$ is a Malament--Hogarth space-time, then
there is a future-directed timelike curve $\gamma_O :[a,b]\rightarrow M$
from a point $q\in J^-(p)$ to $p$ satisfying $\Vert\gamma_O\Vert<\infty$.
The point $q\in M$ can be chosen to lie in the causal future of the past
endpoint of $\gamma_C$. 

Moreover the reason we require fundamental matter fields obeying the
dominant energy condition, geodesically complete initial surfaces etc., is
that we want to exclude the very artificial examples for Malament--Hogarth
space-times.
 
The motivation is the following (for details we refer to \cite{ete-nem}).
Consider a Turing machine realized by a physical computer $C$ moving along
the curve $\gamma_C$ of {\it infinite} proper time. Hence the physical
computer (identified with $\gamma_C$) can perform arbitrarily long
calculations in the ordinary sense. Being $(M,g)$ a Malament--Hogarth
space-time, there is an observer $O$ following the curve $\gamma_O$ 
(hence denoted by $\gamma_O)$ of {\it finite} proper time such that he
touches the Malament--Hogarth event $p\in M$ in finite proper time. But by
definition ${\rm im}\gamma_C\subset J^-(p)$ therefore in $p$ he can
receive the answer for a {\it yes or no question} as the result of an {\it
arbitrarily long} calculation carried out by the physical computer
$\gamma_C$. This is because $\gamma_C$ can send a light beam at arbitrarily 
late proper time to $\gamma_O$. Clearly the pair
$(\gamma_C, \gamma_O)$ is an {\it artificial computing system} i.e., a
generalized computer in the sense of \cite{ete-nem}.

Imagine the following situation as an example. $\gamma_C$ is asked to
check all theorems of our usual set theory (ZFC) in order to check    
consistency of mathematics. This task can be carried out by $\gamma_C$   
since its world line has infinite proper time. If $\gamma_C$ finds a
contradiction, it can send a message (for example a light beam) to
$\gamma_O$. Hence if $\gamma_O$ receives a signal from $\gamma_C$ {\it
before} the Malament--Hogarth event $p$ he can be sure that ZFC set theory
is not consistent. On the other hand, if $\gamma_O$  does not receive any 
signal before $p$ then, {\it after} $p$, $\gamma_O$ can conclude that ZFC
set theory is consistent. Note that $\gamma_O$ having finite proper time
between the events $\gamma_O(a)=q$ (starting with the experiment) and
$\gamma_O(b)=p$ (touching the Malament--Hogarth event), he can be sure 
about the consistency of ZFC set theory in finite (possibly very short)
time. This shows that certain very general formulations of the
Church--Turing thesis cannot be valid in the framework of classical
general relativity \cite{ete-nem}.

One can raise the question if Malament--Hogarth space-times are
relevant or not from a physical viewpoint. We put off this very important
question for a few moments; instead we prove basic properties of
Malament--Hogarth space-times by evoking Lemma 4.1 and Lemma 4.3 from
\cite{ear}. These characteristics are also helpful in looking for
realistic examples. 
\begin{proposition}
Let $(M,g)$ be a Malament--Hogarth space-time with a timelike
curve $\gamma_C$ as above. Then $(M,g)$ is not
globally hyperbolic. Moreover, if $p\in M$ is a Malament--Hogarth event
and $S\subset M$ is a connected spacelike hypersurface such that ${\rm 
im}\:\gamma_C\subset J^+(S)$ then $p$ is on or beyond $H^+(S)$, the future
Cauchy horizon of $S$.
\end{proposition}
{\it Proof.} Consider the point $q\in M$ such that $\gamma_C (0)=q$. 
If $(M,g)$ was globally hyperbolic then $(M,g)$ would be strongly causal 
and in particular $J^-(p)\cap J^+(q)\subset M$ compact. 
We know that im$\:\gamma_C\subset J^-(p)$ hence in fact
im$\:\gamma_C\subset J^-(p)\cap J^+(q)$. Consequently its future
(and of course, past) endpoint are contained in $J^-(p)\cap J^+(q)$ (cf.
Lemma 8.2.1 in \cite{wal}). However $\gamma_C$ is a
causal curve with $\Vert \gamma_C\Vert =\infty$ hence it is future
inextendible i.e., has no future endpoint. But this is impossible hence
$J^-(p)\cap J^+(q)$ cannot be compact or strong causality must be violated
within this set leading us to a contradiction.

Secondly, assume $p\in D^+(S)\setminus\partial D^+(S)$ i.e., $p$ is an
interior point of the future domain of dependence of $S$. Then there is an
$r\in D^+(S)$ chronologically preceded by $p$ (with respect to some time
function assigned to the Cauchy foliation of $D^+(S)$). Letting
$N:=J^-(r)\cap J^+(S)$ then $N\subset D^+(S)$ hence $(N, g\vert_N)$ is a
globally hyperbolic space-time containing the Malament--Hogarth event $p$
and the curve $\gamma_C$. Consequently we can proceed as above to arrive
at a contradiction again. $\Diamond$
\vspace{0.1in}

\noindent The proof of this proposition provides us a characterization of
Malament--Hogarth space-times.
\begin{proposition}
Let $(M,g)$ be a Malament--Hogarth space-time with $p\in
H^+(S)\subset M$ a Mal\-am\-ent--Hogarth event. Consider a timelike
curve $\gamma_C$ as above with {\em im}$\gamma_C\subset J^+(S)$. 
Then either $\overline{J^-(p)\cap S}$ is non-compact or strong causality
is violated at $p\in M$.
\label{mhosztalyozas}
\end{proposition}
{\it Proof.} We just have to repeat the pattern of the proof of the
previous proposition. 

Assume $\overline{J^-(p)\cap S}$ is compact. Then this set is bounded
with respect to $h$ in the geodesically complete $(S,h)$. As we have
seen, there always exists a timelike curve $\gamma_O: [a,b]\rightarrow M$
of finite proper time with $\gamma_O(a)\in S$ and $\gamma_O(b)=p$
(``observer''). Hence by boundedness all points of
$\overline{J^-(p)\cap S}$ can be joined with $p$ by a causal
curve of finite length therefore our assumption implies
that $J^-(p)\cap J^+(S)$ is compact, too. But the complete timelike
curve $\gamma_C: \R^+\rightarrow M$ with $\gamma_C(0)\in S$ is
future inextendible and satisfies im$\gamma_C\subset J^-(p)\cap J^+(S)$
hence there must exist a point $p'\in J^-(p)\cap J^+(S)$ at which strong
causality is violated (again by Lemma 8.2.1 of \cite{wal}). But this
point cannot exists in the interior of $M=D^+(S)$ because this
part is globally hyperbolic therefore must have $p'\in H^+(S)$ and it is a
Malament--Hogarth event in this case as easily seen. Consequently $p=p'$
holds. 

In the opposite way if strong causality is valid in $(M,g)$ then the
same is true for the portion $J^-(p)\cap J^+(S)\subset M$ with 
im$\gamma_C\subset J^-(p)\cap J^+(S)$. But in this case this set cannot be
compact implying $\overline{J^-(p)\cap S}$ is non-compact taking into
account geodesic completeness of $(S,h)$. $\Diamond$
\vspace{0.1in}

\noindent Now we can turn our attention to the existence of physically
relevant examples of space-times containing gravitational computers.
Proposition \ref{mhosztalyozas} indicates that the class of
Malament--Hogarth space-times can be divided into two major subclasses:
the first one contains space-times in which an infinite, non-compact
portion of a spacelike submanifold is visible from some event. The
question is if certain members of these space-times obey some energy
condition or not with some standard matter content (in this case a
space-time is considered as ``physical''). The answer is yes: examples for
such space-times are provided by the following proposition. 
\begin{proposition}
Kerr space-time and the universal covering space of anti-de Sitter
space-time are Malament--Hogarth space-times.
\end{proposition}
{\it Proof.} In the case of the universal cover of anti-de
Sitter space-time (which is a vacuum space with non-vanishing
cosmological constant) this has been proved in \cite{hog1} while the
case of Kerr space-time (this is a vacuum space-time with vanishing
cosmological constant) has been worked out in \cite{ete-nem}. $\Diamond$
\vspace{0.1in}

\noindent (We remark that Reissner--Nordstr\"om space-time, very similar
to the Kerr one, also admits Malament--Hogarth events). The second
subclass consists of those space-times which violate strong causality in a
suitable way. A typical example is presented in the next proposition.
\begin{proposition}
Consider a particular maximal analytical extension of Taub--NUT
space-time. Then this space-time possesses the Malament--Hogarth property
and Malament--Hogarth events are situated along the Cauchy horizons of
this space-time.
\end{proposition}
{\it Proof.} Take the space $S^3\times\R$ equipped with the Taub--NUT
metric $g$ on it (cf. \cite{haw-ell}, pp. 170-178):
\[{\rm d}s^2=-{{\rm d}t^2\over U(t)}+4a^2U(t)({\rm d}\psi
+\cos\Theta{\rm
d}\phi )^2+(t^2+a^2)({\rm d}\Theta^2+\sin^2\Theta{\rm d}\phi^2).\]
This is a spherically symmetric vacuum metric and  
\[ U(t)=-1+{2(mt+a^2)\over a^2+t^2},\:\:\:\:\:m,a\:\:\mbox{are positive
constants}.\]
Moreover $(\phi, \Theta ,\psi )$ are the Euler angles on $S^3$ i.e.,
$0\leq\psi\leq 4\pi$, $0\leq\Theta\leq\pi$ and $0\leq\phi\leq 2\pi$. 
The metric is singular at the zeros of $U$ equal to $t_\pm
=m\pm\sqrt{m^2+a^2}$ hence apparently we have to make the restriction
$t_-<t<t_+$. The time orientation is fixed such that $t$ increases.
However introducing the new variable 
\[\psi ':=\psi +{1\over
2a}\int\limits_{t_0}^t{{\rm d}\tau\over U(\tau )}\:\:\:\mbox{(mod
$4\pi$)},\]
we can rewrite the metric (also denoted by $g$) as
\begin{equation}
{\rm d}s^2=-4a{\rm d}t({\rm d}\psi '+\cos\Theta{\rm d}\phi )+4a^2U(t)({\rm
d}\psi ' +\cos\Theta{\rm d}\phi )^2 
+(t^2+a^2)({\rm d}\Theta^2+\sin^2\Theta{\rm d}\phi^2)
\label{kiterjesztes}
\end{equation}
and now we can allow $-\infty <t<\infty$ without difficulty but the
resulting analytical extension lacks global hyperbolicity. Indeed, the
surfaces given by $t=t_\pm$, diffeomorphic to $S^3$, represent Cauchy
horizons (and coincide if $a=0$). There is another analytical extension of
Taub--NUT space, too. For details see \cite{haw-ell}.

Now we can proceed as follows. Consider a smooth curve $\alpha
:\R^+\rightarrow S^3\times\R$ given by
\begin{equation}
\beta (s):=\left(\Theta_0,\phi_0,\:\: \psi_0+{1\over
2a}\int\limits_0^s{e^{-u}\over U (t(u))}{\rm d}u ,\:\:
t(s)\right),\:\:\:\:\:\mbox{where $t(s)=t_+-e^{-s}$}
\label{betagorbe}
\end{equation}
with respect to the coordinate system (\ref{kiterjesztes}).
The constants $\Theta_0,\phi_0, \psi_0$ are fixed. This curve is clearly
future directed and runs in the region $t_+-1\leq t<t_+$ which is inside
the original globally hyperbolic region $t_-<t<t_+$ if
$2\sqrt{m^2+a^2}>1$. Its tangent vector field is
\[{{\rm d}\alpha\over{\rm d}s}=\dot\alpha (s)=
{e^{-s}\over 2aU(t(s))}{\partial\over\partial\psi
'}+e^{-s}{\partial\over\partial t}\]
of pointwise length $\vert\dot\alpha (s)\vert^2_g =-e^{-2s}/U(t(s))$ 
by using (\ref{kiterjesztes}). But for
$0\leq s<\infty$ that is, for $t_+-1\leq t(s)<t_+$ this is clearly
negative showing that our curve is not only future directed but 
$\vert\dot\alpha (s)\vert^2_g<0$ that is, timelike as well. Its length is
\[\Vert\alpha\Vert=\int\limits_0^\infty{e^{-s}\over
\sqrt{U(t(s))}}\:{\rm d}s\geq\int\limits_0^\infty e^{-s}\:{\rm
d}s=\infty\]
hence is complete. Clearly, $\alpha$ spirals around the Cauchy surface
$t=t_+$ infinitely many times while approaching it.

Furthermore this $\alpha$ can be deformed into a
future directed, timelike complete geodesic, too. To see this, consider
the sequence of points $\{\alpha (s_n)\:\vert\:n\in\N\}$ for a 
suitable monotonly increasing sequence of points $s_n\in\R^+$. Since the
portion of the extended Taub--NUT space-time containing $\alpha$ is
globally hyperbolic and $\alpha (s_n)<<\alpha (s_k)$ if $n<k$ we can find
future directed timelike geodesic segments $\gamma_n$ connecting $\alpha
(s_n)$ with $\alpha (s_{n+1})$. If the partition is dense enough,
$\gamma_n$'s are close to $\alpha_n$'s where $\alpha_n:=\{\alpha
(s)\:\vert\:s\in [s_n,s_{n+1}]\}$. Because geodesics have extremal length 
among curves, we can write for an at least piecewise smooth geodesic
$\gamma_C :\R^+\rightarrow S^3\times\R$ that
\[\Vert\gamma_C\Vert
=\sum\limits_{n\in\N}\Vert\gamma_n\Vert\geq\sum\limits_{n\in\N}\Vert
\alpha_n\Vert=\Vert\alpha\Vert =\infty .\]
In other words our future directed timelike geodesic $\gamma_C$ just
constructed is complete and spirals around the Hopf-circle $S^1:=\{
(\phi_0, \Theta_0,\psi ' , t_+)\:\vert\:0\leq\psi '\leq 4\pi\}$ (here
$\phi_0$ and $\Theta_0$ coincide with constants chosen in
(\ref{betagorbe})). This circle is on the Cauchy horizon $t=t_+$ of the
maximally extended Taub--NUT space-time. Now it is straightforward that at
all the events $p\in S^1$ strong causality is violated by $\gamma_C$ (or
$\alpha$). 

Finally consider an even more simple smooth curve $\beta
:[0,1]\rightarrow S^3\times\R$ which is given with respect to
(\ref{kiterjesztes}) as 
\[\beta (s):=(\Theta_0,\phi_0, \psi_0+ s, t(s)),\:\:\:\:\:\mbox{where
$t(s)=t_+-1+s$}.\]
Hence 
\[\dot\beta (s)={\partial\over\partial\psi '}+{\partial\over\partial t}\]
yielding $\vert\dot\beta (s)\vert^2_g=4a(aU(t(s))-1)$ via
(\ref{kiterjesztes}).
Consequently for $0\leq s\leq 1$ i.e., for $t_+-1\leq t(s)\leq t_+$ one
again has $\vert\dot\beta (s)\vert^2_g<0$ if $a<1$ demonstrating that this
curve is also future directed and timelike of finite length:
\[\Vert\beta\Vert =2\sqrt{a}\int\limits_0^1\sqrt{1-aU(t(s))}\:{\rm
d}s\leq 2\sqrt{a}<\infty .\]
Proceeding as above we can deform this curve into an at least piecewise
smooth timelike geodesic $\gamma_O$ also of finite length. One can see
that $\gamma_C(0)=\gamma_O(0)$ furthermore $\gamma_O$ intersects
$\gamma_C$ infinitely many times hence the point 
$p=\gamma_O(1)=(\Theta_0,\phi_0,\psi_0+1, t_+)\in S^1\subset S^3$ is a
Malament--Hogarth event. Consequently we can interpret $\gamma_C$ as a
``computer'' required in the definition of Malament--Hogarth space-times
moreover $\gamma_O$ as an ``observer''. This shows the Malament--Hogarth
property of Taub--NUT space (together with the technical condition
$0<a<1<2\sqrt{m^2+a^2}$ but this is certainly avoidable by using better
curves). $\Diamond$
\vspace{0.1in}

\noindent After getting the feel of Malament--Hogarth space-times we move
to the next section where their relation with the strong cosmic censorship
is pointed out.

\section{The strong cosmic censor conjecture}

Cosmic censorship is being used to rule out space-times where causality is
violated because of the presence of ``naked singularities''. Since
nowadays we are unable to grasp the notion of a naked singularity in its
full generality we are forced to use some other indirect 
characteristics to remove those space-times where causality
breaks down from the class of physically relevant ones. The most
straightforward notion is global hyperbolicity because in this case all
events can be predicted from an initial data set fixed in advance along a
Cauchy surface. However this restriction is too strong: physically
relevant examples like anti-de Sitter, Taub--NUT 
and even the Kerr--Newman space-times would considered as ``wrong'' by
requiring simply global hyperbolicity. Consequently the recent
versions of the strong cosmic censor conjecture are formulated by
postulating global hyperbolicity for ``physically relevant''
space-times but also providing a ``list'' of
sporadic examples lacking global hyperbolicity nevertheless considered as 
``physically relevant''.

A more or less up-to-date formulation is given in \cite{wal} on p. 305.
\begin{conjecture} {\em (standard formulation of the strong cosmic censor
conjecture)} Let $(S,h,k)$ be an initial data set for Einstein's equation,
with $(S,h)$ a complete Riemannian manifold. Suppose a fundamental matter
field is given represented by its stress-energy tensor $T$ satisfying the
dominant energy condition. Then, if the maximal Cauchy development of this
initial data set is extendible, for each $p\in H^+(S)$ in any extension,
either strong causality is violated at $p$ or $\overline{J^-(p)\cap S}$ is
non-compact.
\end{conjecture}
It is quite surprising that in light of Proposition \ref{mhosztalyozas} 
Malament--Hogarth events have exactly the same properties as points
have on the Cauchy horizons in the above formulation. In other words
the content of Proposition \ref{mhosztalyozas} is that Malament--Hogarth
property implies the behaviour for the Cauchy horizon required
by the strong cosmic censor conjecture.

To see precise equivalence we should establish a converse to Proposition
\ref{mhosztalyozas}. That is, if a space-time is non-globally hyperbolic
and for its events $p\in H^+(S)$ either $\overline{J^-(p)\cap S}$ is
non-compact or strong causality is violated at $p$ then is $(M,g)$ 
a Malament--Hogarth space-time? The answer is certainly no 
because the conditions are insufficient. For instance we should know
something on the length of the Cauchy development in question.

However the statement ``for each $p\in H^+(S)$ in any extension,  
either strong causality is violated at $p$ or $\overline{J^-(p)\cap S}$ is
non-compact'' is being used only to incorporate the 
classical examples like Kerr--Newman or Taub--NUT into the allowed class
of space-times. But we have seen that these examples possess the
Malament--Hogarth property. In other words the Malament--Hogarth property
appears as a unifying way to enumerate the important and known sporadic
examples in the strong cosmic censor conjecture considered as ``still
physically relevant'' although they are non-globally hyperbolic i.e.,
which posses an initial value formulation but their maximal Cauchy
developments are extendible. Hence we cannot resist the temptation to
reformulate the above conjecture as follows.
\begin{conjecture} {\em (reformulation of the strong cosmic censor
conjecture)} Let $(S,h,k)$ be an initial data set for Einstein's equation,
with $(S,h)$ a complete Riemannian manifold. Suppose a fundamental matter
field is given represented by its stress-energy tensor $T$ satisfying the
dominant energy condition. Then, if the maximal Cauchy development of this
initial data set is extendible, this extension is a Malament--Hogarth
space-time.
\end{conjecture}
We can examine the strong cosmic censor scenario from a naive stability
point of view as well. It is well-known (cf. Theorem 8.3.14 of \cite{wal})
that a globally hyperbolic space-time is stable causal. But stable
causality is a stable property under small perturbations of
the metric (here ``small'' is understood as follows: if $g$ is the
original metric and $g'=g+h$ is its perturbation then for the tensor
field $h$ we have $\Vert h\Vert$ small in an appropriate Sobolev norm
on the space of metrics over a fixed manifold $M$). Consequently global
hyperbolicity seems to be a stable property hence globally hyperbolic
space-times apparently form the ``interior'' of the set of ``physically
reasonable'' space-times allowed by the strong cosmic censor.

On the other hand, it is known or at least conjectured that
space-times like (maximal extensions of ) Kerr or Taub--NUT are unstable
in the sense that small perturbations of these metrics deform the Cauchy
horizons of these space-times into a real curvature singularity. In the
case of black holes this is called mass inflation (e.g. cf.
\cite{poi-isr}). Therefore these space-times may be considered as unstable
in an appropriate fashion. Or, turning the coin, we can intuitively say
that the Malament--Hogarth property is an unstable property of
``physically relevant'' space-times and form the ``borderline'' between
allowed and non-allowed space-times by the strong cosmic censorship. It is
quite interesting that exactly this unstable thin class admits
gravitational computers providing non-Turing computations.

The whole conjectured situation is sketched in the following figure.
\vspace{0.1in}

\centerline{\psfig{figure=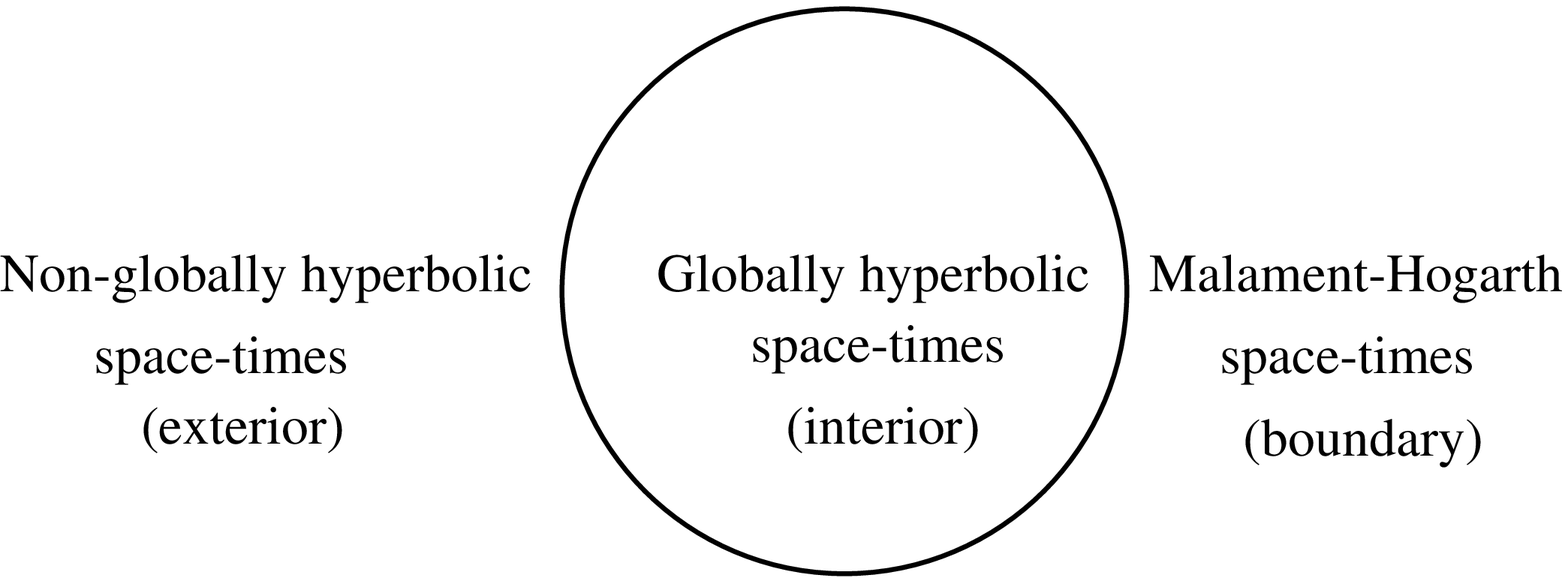,height=1.8in}}
\vspace{0.1in}

\centerline{{\bf Figure 1.} The strong cosmic censorship scenario}
\section{Concluding remarks}
As a conclusion we have to emphasize again that our considerations require
future work for example it is important to understand if other marginal
space-times (from the point of view of the strong cosmic censor) admit the
Malament--Hogarth property or not.

Another important problem is to understand the stability properties of
Malament--Hogarth space-times mentioned above and see if
these indeed form a kind of boundary in a strict
functional analytic sense for physically relevant metrics over a
fixed manifold.
\vspace{0.1in}

\noindent{\bf Acknowledgement.} The author is grateful to I. N\'emeti, H.
Andr\'eka, G. S\'agi (R\'enyi Institute, Hungary) and M. Hogarth (Univ. of
Cambridge, UK) for the useful discussions.

\end{document}